\documentclass[twocolumn]{aastex631}
\usepackage[T1]{fontenc}
\usepackage{float}
\usepackage{longtable}
\usepackage{multirow}
\usepackage{graphicx}
\usepackage{tabularx}
\usepackage{lipsum}
\usepackage{booktabs}
\DeclareUnicodeCharacter{2212}{-}
\newcommand{\psr}{PSR~J1242$-$4712}

\begin{document}

\title{The GMRT High-Resolution Southern Sky Survey for pulsars and transients -- VII: Timing of Spider MSP J1242$−$4712, A Bridge Between Redback and Black Widow Pulsars}

\author[0009-0002-3211-4865]{Ankita Ghosh}
\affiliation{National Centre For Radio Astrophysics, Tata Institute of Fundamental Research, Pune 411007, India}

\author[0000-0002-6287-6900]{Bhaswati Bhattacharyya}
\affiliation{National Centre For Radio Astrophysics, Tata Institute of Fundamental Research, Pune 411007, India}

\author{Andrew Lyne}
\affiliation{Jodrell Bank Centre for Astrophysics, School of Physics and Astronomy, The University of Manchester,
Manchester M13 9PL, UK}

\author[0000-0001-6295-2881]{David L.\ Kaplan}
\affiliation{Center for Gravitation, Cosmology, and Astrophysics, Department of Physics, University of Wisconsin-Milwaukee, PO Box 413,
Milwaukee, WI, 53201, USA}

\author[0000-0002-2892-8025]{Jayanta Roy}
\affiliation{National Centre For Radio Astrophysics, Tata Institute of Fundamental Research, Pune 411007, India}

\author[0000-0002-5297-5278]{Paul S. Ray}
\affil{Space Science Division, U.S. Naval Research Laboratory, Washington, DC 20375-5352, USA}

\author[0000-0001-9242-7041]{Ben Stappers}
\affiliation{Jodrell Bank Centre for Astrophysics, School of Physics and Astronomy, The University of Manchester, Manchester M13 9PL, UK}

\author[0000-0002-3764-9204]{Sangita Kumari}
\affiliation{National Centre For Radio Astrophysics, Tata Institute of Fundamental Research, Pune 411007, India}

\author{Shubham Singh}
\affiliation{National Centre For Radio Astrophysics, Tata Institute of Fundamental Research, Pune 411007, India}

\author[0009-0001-9428-6235]{Rahul Sharan}
\affiliation{National Centre For Radio Astrophysics, Tata Institute of Fundamental Research, Pune 411007, India}

\begin{abstract}

We present the timing solution for the 5.31-ms spider millisecond pulsar (MSP) J1242$-$4712, discovered with the GMRT. PSR J1242$-$4712 orbits a companion of minimum mass 0.08 M$_{\odot}$ with an orbital period of 7.7 hrs and occupies a relatively unexplored region in the orbital period versus companion mass space. We did not detect gamma$-$ray pulsations for this MSP, and also could not identify the optical counterpart for PSR J1242$−$4712 in available optical/near-infrared data. The profile of J1242$-$4712 evolves with frequency showing a clear single component at lower frequencies and a three-component profile at 650 MHz.  \psr\ eclipses for a very short duration near superior conjunction (orbital phase $\sim$ 0.23$-$0.25) below 360 MHz. Moreover, significant DM delays and errors in pulse times of arrivals are observed near inferior conjunction (orbital phase $\sim$ 0.7), along with an observed eclipse in one epoch at 650 MHz. Observed eclipses and significant orbital period variability suggest that PSR J1242$-$4712 is possibly not a He$-$WD binary, but has a semi- or non-degenerate companion, indicating that this is a ``spider'' MSP lying in a region between typical black widows and redbacks. This system may represent a distinct category of spider MSPs, displaying characteristics that bridge the gap between known black widow and redback MSPs.

\end{abstract}

\keywords{Pulsars (53) --- Millisecond Pulsars(60) --- Binary Pulsars(16)---Spider pulsars(24)}

\section{Introduction} \label{sec:intro}

The leading theory for the origin of millisecond pulsars (MSPs) is that they are rapidly rotating neutron stars that have been ``recycled" or spun up through the accretion of matter from a companion in a close binary system \citep{1982Natur.300..728A}. It is thought that in most cases, the end product of this mass transfer is an MSP binary with a low-mass white dwarf companion in a relatively wide orbit \citep{2006csxs.book..623T}. 
MSP binaries in the Galactic Field serve as valuable indicators of binary evolution, as they are less likely to undergo dynamic exchange encounters compared to systems in globular clusters.
After accretion, the intense pulsar wind can wear away the companion star  \citep{1988Natur.334..225K,1988Natur.333..832P,1988Natur.334..227V} resulting in the creation of MSP binaries with very low-mass companions. These MSPs with mostly hydrogen-rich, non-degenerate companions in compact binary orbits (orbital periods $<1\,$d) are classified as ``spider MSPs'', and are further divided by companion mass $M_c$ into  ``black widow (BW)'' ($M_{c}<0.05\,M_{\odot}$) and ``redback (RB)'' ($0.1\,M_{\odot} < M_{c} <0.9\,M_{\odot}$) spiders \citep{https://ui.adsabs.harvard.edu/link_gateway/2013IAUS..291..127R/ADS_PDF, 2012arXiv1205.3089R}. 
Only 20\% of the total MSP population can be classified as ``spider'' MSPs{\footnote{\label{note2}\url{https://www.atnf.csiro.au/research/pulsar/psrcat/}  as of Aug 17, 2023}} \citep{2005AJ....129.1993M}.\par
In these compact systems, the highly energetic wind from the pulsar ablates the companion, leaving ionized material in the orbit which causes an eclipse of the pulsar's radio emission. 
Such eclipsing MSP systems can aid in the understanding of properties of the low-mass companions in tight binary orbits, plasma properties of eclipse material, mass flow from the companion driven by relativistic pulsar wind, orbital properties in strong gravitational potential. Black widow and redback systems offer excellent opportunities for measuring the masses of neutron stars. The neutron star masses in black widow and redback systems that have been measured so far tend to exceed $1.4\,M_{\odot}$ \citep{2012ApJ...760L..36R,2013ApJ...776...20C,2016ApJ...833..138R,2019ApJ...872...42S,2022ApJ...934L..17R}, sometimes by significant factors, which agrees with theoretical expectations \citep{1994ARA&A..32..591P,2014ApJ...786L...7B} and allows powerful probes of the neutron stars mass distribution and also contributes valuable information to studies concerning the equation of state of neutron stars \citep{2012ARNPS..62..485L}. The discovery of redback systems transitioning between accretion-powered low-mass X-ray binary and rotation-powered radio pulsar states, often occurring within a few years, further supports the recycling hypothesis, potentially influenced by irradiation feedback and accretion disk instabilities \citep{2014ApJ...790...39S,2015MNRAS.449.4184B}.
The measurement of pulsar masses in these systems, therefore, contributes significantly to a better understanding of the recycling process. \par

PSR J1242$-$4712 is a 5.31-ms MSP, with a dispersion measure (DM) of 78.6 pc cm$^{-3}$, discovered with the GMRT High-Resolution Southern Sky Survey\footnote{\url{http://www.ncra.tifr.res.in/~bhaswati/GHRSS.html}} \citep{https://doi.org/10.3847/0004-637X/817/2/130, https://doi.org/10.3847/1538-4357/ab2bf3} by Bhattacharyya et al. (in prep), in the 80-arcmin-diameter beam of the Incoherent array (IA) pointing toward 12$^\mathrm{h}$43$^\mathrm{m}$00$^\mathrm{s}$, $-$47\degr00\arcmin00\arcsec. 
GMRT is a radio interferometric array that is comprised of 30 dishes, each 45 meters in diameter. These dishes are distributed in a Y-shaped array across a 25 km-wide baseline and operate across a range of five frequencies, spanning from 150 MHz to 1450 MHz \citep{1991ASPC...19..376S, 2017CSci..113..707G}. The GMRT is well suited for pulsar timing observations due to its high sensitivity and wide frequency coverage. 

Also, as an interferometer, GMRT offers us an opportunity to localize the pulsar positions with good accuracy and thus permit faster convergence in achieving timing models.
Recently, this MSP has been localized with a positional accuracy of $1\arcsec$ as reported in \citet{https://dx.doi.org/10.3847/1538-4357/acc10f} using continuum imaging followed by multiple Phased Array (PA) beamforming \citep{2012MNRAS.427L..90R}.\par

In Section \ref{sec:Observations} we present the observations using upgraded GMRT (uGMRT) spanning nearly 5 years. In Section \ref{sec:Timing study} we detail the timing study of the system and in Section \ref{sec:Multiwavelength detection} we discuss the detectability of any multi-wavelength counter-part of the MSP  In Section \ref{sec:Profile Evolution} we discuss the frequency-dependent profile evolution of the system.  Finally, in Section \ref{sec:Discussion} we discuss the results we obtained, and summarize our conclusions in Section \ref{sec:Conclusion}.

\section{Observations and Data Analysis} \label{sec:Observations}

After its discovery, we  regularly followed up PSR J1242$-$4712 at band-3 (300$-$500 MHz), band-4 (550$-$850 MHz), and band-5 (1050$-$1450 MHz) using observations from the uGMRT \citep{1991ASPC...19..376S, 2017CSci..113..707G, http://dx.doi.org/10.1142/S2251171716410117}.
The initial few epochs of follow-up observations showed rapid drift of the pulses, indicating that this MSP is in a tight binary orbit with an orbital period of only a few hours, therefore, frequent observations were taken using the IA mode of the GMRT to more densely sample the orbit. After localization by \citet{https://dx.doi.org/10.3847/1538-4357/acc10f}, we followed up the MSP with the sensitive phased array (PA) beam in GWB band-3, band-4, and band-5. The details of the follow-up IA as well as PA observations are given in Table \ref{tab:1}.\par

\renewcommand{\tabcolsep}{2.5pt}
\begin{table}
\begin{center}
\caption{Summary of observations}
\footnotesize{
\label{tab:1}
\begin{tabular}{cccccccc}
\toprule
Receiver  &  Frequency & Mode{\color{blue}$^{a}$} & T$_{res}${\color{blue}$^{b}$} &  N$_{ch}${\color{blue}$^{c}$} & $\sigma_{TOA}${\color{blue}$^{d}$} & No of  &  Span \\ 
band & (MHz) &  & ($\mu$s) & &($\mu$s)  &epochs & \\
\hline
 \multirow{2}{*}{Band 3}  & \multirow{2}{*}{300$-$500} & IA & 81.92 & 4096 & 25 & 56 & 2019$-$2021 \\

                        &                              & CD & 10.24 & 512 & 0.44 & 9 & 2022$-$2023 \\
 \hline
 Band 4  & 550$-$750 & CD & 10.24 & 512  & 0.51 & 5 & 2022-2023  \\
                       
                         \midrule
Band 5$^{\dagger}$                   & 1050$-$1450                & PA & 81.92 & 4096 & 1.31 & 1  & 2022                                                 \\ \bottomrule
 \hline
\end{tabular}
}
\end{center}
{\footnotesize {\bf{Notes.}}\\
$^{a}$IA: Incoherent Array observations.\\ 
$^{a}$CD: Phased Array observations with coherent de-dispersion.\\
$^{b}$Time resolution\\
$^{c}$Number of channels\\
$^{d}$ToAs are calculated for each mode of observation of an average duration of 45 minutes in band 3, band 4, and band 5.\\
$^{\dagger}$Band-5 is used only for profile study due to poor signal-to-noise ratio.}

\end{table}

Initial IA (FWHM of $80\arcmin$ at 322 MHz) observations from 2019 to 2021 were performed at the central frequency of 400 MHz with the uGMRT using 200 MHz bandwidth. In follow-up observations from 2022, the PA beams in band-3 and band-4 were recorded with online coherent dedispersion (CD), in which the spectral voltages in each frequency sub-band are corrected for dispersive delays using the nominal value of DM. We recorded Stokes-I data at a rate of 48 MB/s, employing 8-bit samples. For PA observations with incoherent dedispersion, we used 4096 channels with a sampling interval of 81.92 $\mu$s, while for CD observations, we used 512 channels with a sampling time of 10.24 $\mu$s.  
To effectively mitigate the impact of narrowband and short-duration broadband RFI, we used radio frequency interference (RFI) mitigation software in conjunction with the GMRT pulsar tool (gptool\footnote{\url{https://github.com/chowdhuryaditya/gptool}}).
In order to eliminate inter-channel smearing, we performed incoherent dedispersion on both the filterbanks obtained from coherent and incoherent dedispersion modes.\par
We performed incoherent de-dispersion \citep{2004hpa..book.....L} at an assumed DM of 78.62 pc cm$^{-3}$ for the MSP using the {\sc{PRESTO}} \citep{2002AJ....124.1788R} tasks ``{\sc{prepdata}}'' and ``{\sc{prepsubband}}''. We then conducted a search for pulsar periodicity by Fourier transforming the de-dispersed time series using ``{\sc{realfft}}'' of {\sc{PRESTO}}. Subsequently, we used the ``{\sc{accelsearch}}'' task of {\sc{PRESTO}} to find pulsar acceleration. Then to obtain the folded profile from each epoch's observations, we used the {\sc{PRESTO}} task ``{\sc{prepfold}}'', to fold each data set using the candidate's observed period and period derivative.\par
Thereafter, we calculated the times of arrival (TOAs) of pulses from each folded profile using the get\_TOAs.py module of {\sc{PRESTO}}. Each folded profile was cross-correlated with a template profile, which was a high signal-to-noise profile from a previous observation, to determine the temporal shift between these two profiles and to obtain TOAs. 

\section{Timing Study} \label{sec:Timing study}

\begin{figure*}
    \plottwo{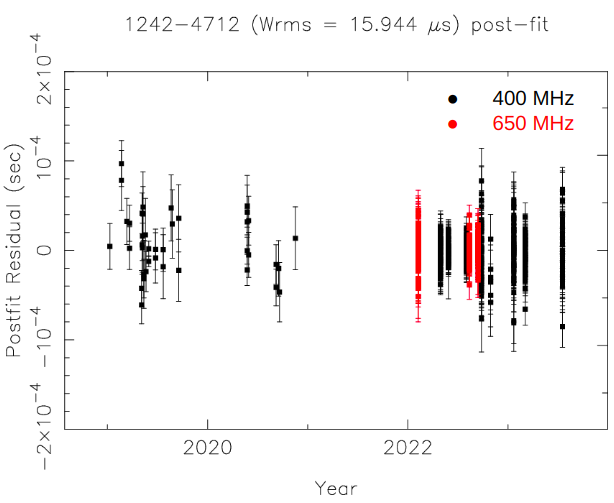}{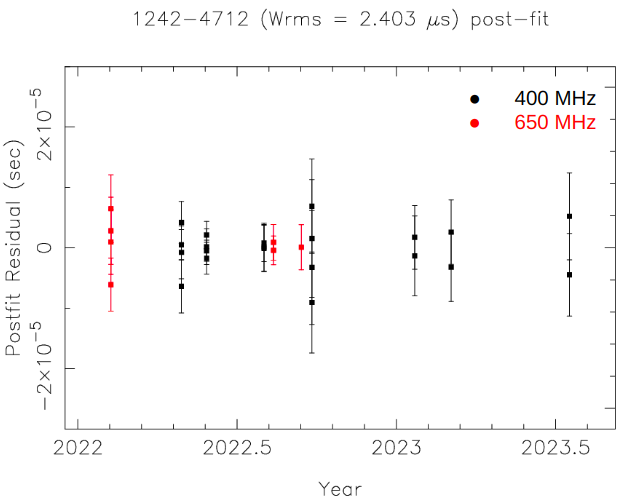}
    \caption{Timing residuals for PSR J1242$-$4712 as a function of time. Left Panel: Post$-$fit timing residual from 4.5 years of timing for PSR J1242$-$4712 using IA and CD observations at uGMRT band-3 and band-4. Right Panel: Post-fit timing residual from 1.5 year of timing for PSR J1242$-$4712 using CD observations at uGMRT band-3 and band-4.}
    \label{fig:1}
\end{figure*}

Pulsar timing involves a repetitive process of fitting a model based on trial values of rotational, astrometric, and binary parameters to the TOAs, finally yielding a timing solution. 
We conduct a least-squares fitting process for the diverse pulsar parameters, achieved by minimizing the sum of the squared timing residuals, the differences between actual times of arrival and the model values. 
As PSR J1242$-$4712 is in a compact binary system, drifts in the pulse phase were evident in the folded data due to unmodelled binary motion.
We estimated the variation of the observed spin period by fitting a period to the TOAs of pulses during the span of observations for each epoch. Then we used fitorbit\footnote{\url{https://github.com/vivekvenkris/fitorbit}} to fit the variation of the observed spin period for multiple observing epochs with binary and astrometric parameters to obtain an initial timing model. \par
The parameters from this incoherent solution were used as input parameters for obtaining a phase coherent solution for the system using the {\sc{TEMPO2}} pulsar timing package \citep{2006MNRAS.369..655H}. 
The phase connection step is where the precise rotation numbers between the arrival times are definitively determined, leading to the refinement of the timing solution through the optimization of pulsar parameters. Using the BTX binary model (\citealt{1976ApJ...205..580B}, D. J. Nice, unpublished) in {\sc{TEMPO2}} to perform the timing analysis, we obtained the phase-connected solution using our initial timing solution and closely spaced TOAs as input and iteratively included TOAs from further observations. Using the obtained parameter file, we refolded the data from each observing epoch. This improved the detection significance by removing the remaining drift in the pulse phase caused by unmodelled binary motion. \par
For the final fit, in order to minimize the effects of any eclipses on the timing, TOAs associated with the potential eclipse regions (orbital phases around 0.18$–$0.25 and 0.68$-$0.73 - see Section \ref{sec:Eclipse}) were eliminated from the analysis.
A comprehensive timing solution was achieved by using all observations (IA and CD). The resulting timing model is presented in Table \ref{tab:2} and yields a rms residual of 15.9 $\mu$s. The timing residuals are shown in Figure \ref{fig:1}.\par
This 5.31-ms MSP is found to be in a 7.7-hour-period orbit with a low-mass companion. As seen for similar compact orbit systems, the eccentricity is very low and is found to be 0.00010(2). The observed spin period derivative indicates a spin-down energy loss rate $\mathrm{\dot{E}}$ of around 11.6 $\times 10^{34}$ erg s$^{-1}$ ($\dot{E}  \equiv 4\pi^{2}I\dot{P}P^{-3}$, where $I$ is the assumed moment of inertia of neutron star, $10^{45}$ g cm$^{2}$). 
The characteristic age of the pulsar is $\tau_{c}$  $\equiv$ $P/2\dot{P}$ $\sim$ 3.9 Gyr and the estimated surface magnetic field of the neutron star, $B_{0}$ $\equiv$ 3.2$\times$ $10^{19}$ $(P\dot{P})^{1/2}$ is 3.4 $\times$ $10^{8}$ G. These values align with the typical findings for MSPs. The derived minimum (for an assumed orbital inclination of 90$^{\circ}$) and median (for inclination 60$^{\circ}$) companion masses are 0.0872 M$_{\odot}$ and 0.1015 M$_{\odot}$ respectively for an assumed pulsar mass of 1.4 M$_{\odot}$.  We also obtain a precise value for the DM of 78.636(1) pc cm$^{-3}$.  In order to accommodate an observed drift in DM, we incorporated the first derivative of DM into the timing model and obtained a significant value of 0.0081(7) pc cm$^{-3}$yr$^{-1}$.  We also report a significant detection of orbital period derivative (orbital period 1st derivative $\sim$ $-1.16(1)$ $\times$ $10^{-11}$ (s/s)) for this system, which is similar to that of other spider MSP systems \citep{https://doi.org/10.1093/mnras/stw1737, https://doi.org/10.1093/mnras/stx1533} and can indicate changes in the companion's gravitational quadrupole moment \citep{https://doi.org/10.1111/j.1365-2966.2011.18610.x, https://ui.adsabs.harvard.edu/link_gateway/1994ApJ...436..312A/doi:10.1086/174906}.
The measured $\dot{E}$, short orbital period, and orbital period derivative indicate strong tidal interactions between the pulsar and its companion. It's worth noting that no significant trends were observed in the remaining timing residuals. However, by incorporating more TOAs from sensitive CD observations, the timing model can remove residual trends better. Given the current precision of timing and the data set at present, the ephemeris provided in Table \ref{tab:2} accurately describes the present TOAs. \par
Finally, precision timing was conducted using only CD observations from bands 3 and 4, resulting in a rms residual of 2.4 $\mu$s. For the timing of more sensitive CD observations, TOAs with uncertainties less than 10 $\mu$s were exclusively considered. The timing residual plot exclusively from CD observations is presented in the right panel (Figure \ref{fig:1}).\par

\begin{table*}
\begin{center}
{\footnotesize
\caption{Timing parameters for PSR J1242$-$4712.
\label{tab:2}}
\begin{tabular}{lc}
\hline
Parameters & Values  \\
\hline
 Timing Data Span     \dotfill     & 58491.9$-$60142.5 \\
 Period epoch (MJD)\dotfill          &  59126.79 \\
 Total time span (year) \dotfill       & 4.5   \\
 Number of TOAs\dotfill  & 587  \\
 Post-fit residual rms ($\mu$s)\dotfill       & 15.944    \\
 \hline
 Right ascension (J2000)\dotfill &
 12$^\mathrm{h}$42$^\mathrm{m}$12\fs7954(1)  \\
 Declination (J2000)\dotfill     &
 $-$47\degr12\arcmin18\farcs408(1)  \\
 Spin frequency $f$ (Hz)\dotfill &
 188.2052895367(1) \\
 Spin frequency derivative $\mathrm{\dot{f}}$ (Hz s$\mathrm{^{-1}}$)\dotfill &
 $-$ 7.596(1)$\times$10$^{-16}$  \\
2nd Spin frequency derivative $\mathrm{\ddot{f}}$ (Hz s$\mathrm{^{-2}}$)\dotfill &
 $-$5.0(4)$\times$10$^{-25}$  \\
 Dispersion measure $\mbox{DM}$ (pc~cm$^{-3}$) &  78.636(1)      \\
 DM $1^{\mathrm st}$derivative $\mbox{DM1}$ (pc~cm$^{-3}$ year$^{-1}$) \dotfill & 0.0081(7)   \\
 Orbital frequency, $f_{orb}$ $\mbox{Hz}$ \dotfill  &  3.57418936(9) $\times$ $10^{-5}$\\
 Orbital frequency 1st derivative (s/s)\dotfill & $-$1.4(1) $\times$ $10^{-20}$  \\
 Projected semi-major axis $\mathrm{x}$ (lt-s) &  0.315643(1)  \\
Eccentricity (e) \dotfill & 0.00010(2)  \\
Epoch of periastron $\mathrm{T_{\ o}}$ (MJD) \dotfill &  58613.5767300(8) \\
Planetary Ephemeris                & DE421 \\
Time units                         & TDB \\
 \hline
 \multicolumn{2}{c}{Derived parameters} \\
  \hline
Galactic longitude, $\mathrm{l}$ (\degr) \dotfill    & 301.31  \\
Galactic latitude, $\mathrm{b}$ (\degr) \dotfill    & 15.63   \\
DM distance (kpc) \dotfill              & 2.48 $^\dagger$       \\
 & 2.51$^\ddagger$ \\
Orbital period $\mathrm{P_{b}}$ (days)\dotfill &  0.3238237507(8)  \\
Mass function $\mathrm{M_{\odot}}$\dotfill &    0.000321998(2) \\
Companion mass, $\mathrm{M_c}$  (M$_{\odot}$)      \dotfill                           &  0.0872 $<$ 0.1015 $<$ 0.2117 \\ 
Observed spin period derivative, $\mathrm{\dot{P}_{\rm obs}}$ ($\mathrm{s\, s^{-1}}$) \dotfill                             & 2.1446(5) $\times 10^{-20}$  \\
orbital period 1st derivative, $\mathrm{\dot{P}_{\rm b}}$  ($\mathrm{s\, s^{-1}}$) \dotfill                             & 1.16(1) $\times 10^{-11}$  \\
Energy loss rate $\mathrm{\dot{E}}$ ($10^{34} \rm \, erg \, s^{-1}$)\dotfill                   &  11.6       \\
Characteristic age, $\mathrm{\tau_c}$ (Gyear)\dotfill                &  3.928          \\
Surface magnetic field, $\mathrm{B_0}$ (Gauss)\dotfill      &  $3.416 \times 10^{8}$        \\
Uncalibrated flux density at 400 MHz, $S_{400}$ $\mbox{(mJy)}$     & 0.5 \\ \hline
\end{tabular}
}
\end{center}

{\footnotesize {\bf{Notes.}}\\
$^\ast$ Errors correspond to 1$\sigma$.\\
$^\dagger$ using the NE2001 model electron density distribution\\
$^\ddagger$ using the YMW16 model of electron distribution\\
We note that the calculated DM distance is model-dependent.
The numbers in the parenthesis are uncertainties in the preceding digits.}
\end{table*}

\section{Multiwavelength Searches} \label{sec:Multiwavelength detection}

\subsection{$\gamma$-ray Pulsations}\label{sec:gamma ray}
There is no gamma-ray source in the Fermi-LAT 14-year (4FGL-DR4) catalog \citep{https://doi.org/10.48550/arXiv.2307.12546} associated with this MSP. Therefore, if there is an associated gamma-ray source, it is very faint. We searched for gamma-ray pulsations by folding the LAT photons using the timing model parameters from the radio timing observations covering approximately 4.5 years. We applied `simple' photon weights scanning over 6 trials as described by \citet{2019A&A...622A.108B}. No significant pulsed signal was seen. 

We used the 3PC sensitivity map \citep{https://doi.org/10.3847/1538-4357/acee67} to compute a limit of the minimum detectable integral energy flux $G_{100}$ $\sim$ 6.2 $\times$ $10^{-13}$ erg/s/cm$^{2}$, so if the pulsar emits gamma rays, the flux is below that level. The gamma ray luminosity is defined as, $L_{\gamma}$=$4\pi d^{2} f_{\Omega} G_{100}$; where d is the distance to the pulsar and f$_{\Omega}$ is the beaming fraction. 
The factor f$_{\Omega}$ is the ratio of total beam power averaged over the whole sky to the power in the beam slice illuminating the Earth averaged in phase. It depends on both the inclination between the neutron star's magnetic and rotation axes and the inclination between the rotation axis and the line of sight \citep{http://dx.doi.org/10.1088/0067-0049/208/2/17}. At a distance of 2.5 kpc, the minimum detectable integral energy flux (G$_{100}$) corresponds to a luminosity of 4.6 $\times$ $10^{32}$ erg/s (assuming $f_{\Omega}$=1), which is 0.4\% of pulsar spin-down power ($\dot{E}$).
This places \psr\ at a very low efficiency relative to other MSPs in the gamma-ray efficiency ($\eta$ = $L_{\gamma}$/$\dot{E}$)  vs. spin-down power ($\dot{E}$) plotted in Figure 24 of \cite{https://doi.org/10.3847/1538-4357/acee67}. Thus the pulsar could either be intrinsically inefficient or $f_{\Omega}$ for this pulsar is quite large, potentially due to a face-on viewing angle, so much of the gamma-ray flux is not hitting the Earth.

\subsection{X-ray and optical observations}\label{sec:X-ray and optical}

\begin{figure*}
\plotone{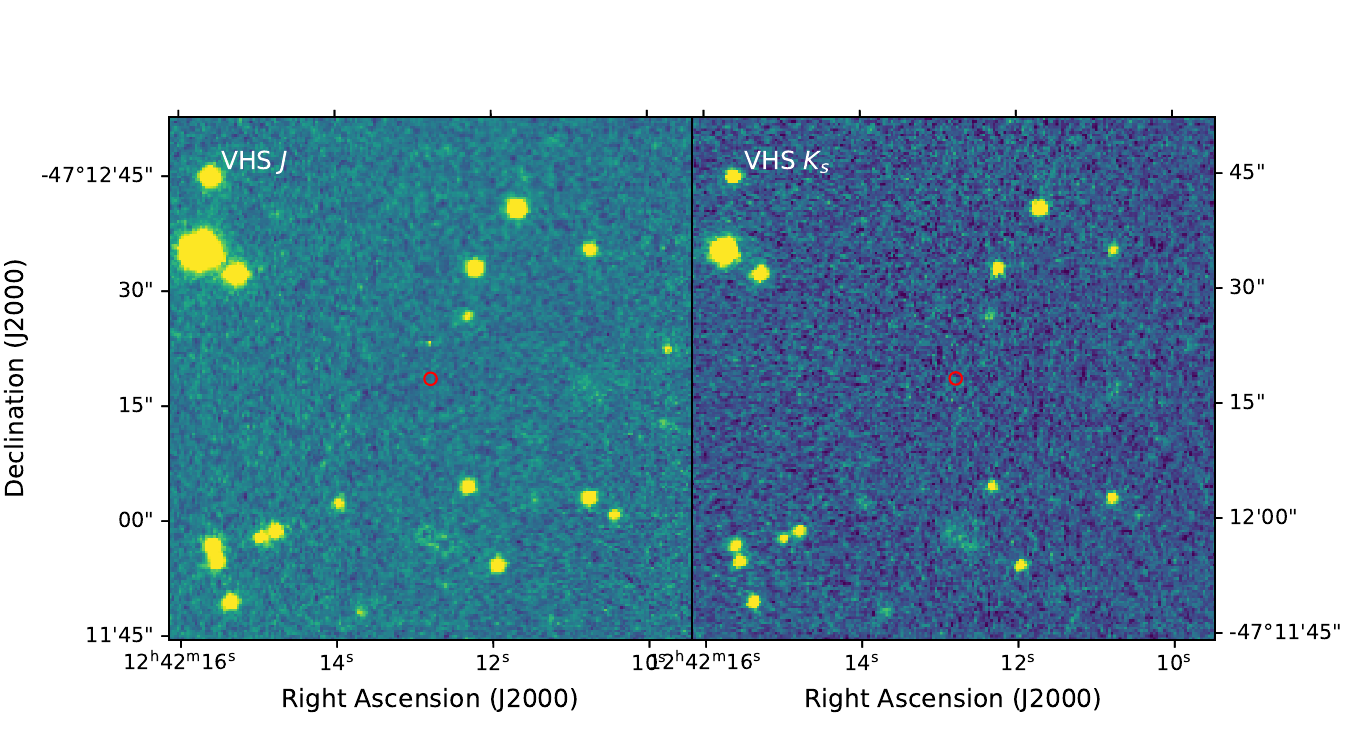}
\caption{VHS $J$ (1.2\,$\mu$m, left) and $K_s$ (2.1\,$\mu$m, right) near-infrared  images of the field of \psr.  The position of \psr\ is marked with a red circle.}
\label{fig:image}
\end{figure*}

We have searched for X-ray counterparts using the radio timing position in existing \textit{Chandra}, \textit{XMM-Newton}, or \textit{Neil Gehrels Swift Observatory} archival data\footnote{\url{https://cxcfps.cfa.harvard.edu/cda/footprint/cdaview.html}}. No X-ray observation is available.
We searched for available optical/near-infrared surveys to constrain the counterpart to \psr.  The deepest we found were SkyMapper \citep{2018PASA...35...10W,2019PASA...36...33O} in the optical, and the Vista Hemisphere Survey (VHS; \citep{2013Msngr.154...35M}) in the near-infrared.  We show the $J$ (1.2\,$\mu$m) and $K_s$ ($2.1\,\mu$m) images from VHS in Figure~\ref{fig:image}.  No source is visible within the sub-arcsecond error circle of \psr\ in any of the available data.  We set 5$\sigma$ limits of $g=20.1$, $r=20.6$, $i=21.3$ (AB) in the SkyMapper data based on the noise levels in nearby sources, and $J=20.4$ and $K_s=19.7$ (AB) in the VHS data based on the \texttt{ABMAGLIM} values in the data headers.\par

\begin{figure}
\centering
\includegraphics[width=1.0\linewidth]{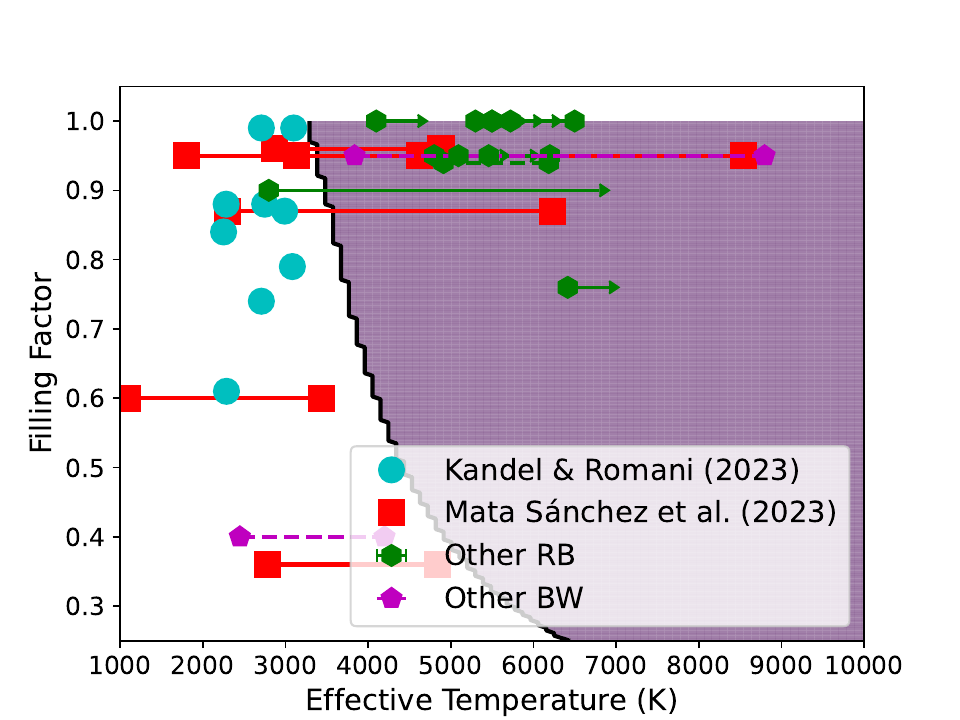}
\caption{Constraints on the companion to \psr, based on available data from VHS and SkyMapper. The filled purple region shows the excluded region as a function of effective temperature and Roche-lobe filling factor, assuming a median inclination of $60\degr$ and a neutron star mass of $1.4\,M_\odot$. We also plot black widow sources from \citet{https://doi.org/10.3847/1538-4357/aca524} (cyan circles) and \citet{https://doi.org/10.1093/mnras/stad203} (red squares, where the lines go from the night-side temperatures to the day-side temperatures).  Individual black widow (magenta pentagons) and redbacks (green hexagons) from the literature are also shown \citep{2013ApJ...769..108B,2015ApJ...804L..12S,2019ApJ...872...42S,2015MNRAS.454.2190D,2017ApJ...849...21B,2017MNRAS.472.4287S,2017ApJ...851...31S,2019ApJ...876....8S,2022ApJ...941..199S,2018MNRAS.476.1086R,2023MNRAS.525.2565T}.
}
\label{fig:limits}
\end{figure}

We then used the limiting magnitudes of these data sets to constrain the companion. We assumed a nominal distance of 2.5\,kpc (based on the pulsar's dispersion measure; \citealt{2002astro.ph..7156C,2017ApJ...835...29Y}) and an optical extinction of $A_V=0.5$\,mag (assuming the maximum extinction from \citealt{1998ApJ...500..525S}, although scaling from the pulsar's dispersion measure using \citealt{2013ApJ...768...64H,1995A&A...293..889P} gives a value closer to 1.3).  We varied the effective temperature and Roche-lobe filling fraction to predict the optical/near-IR magnitudes and compared them with the upper limits.  The current limits (Figure~\ref{fig:limits}) largely exclude effective temperatures $>6000\,$K, and down to about $4000\,$K for filling fractions approaching 1.  However, these are not enough to exclude the range of typical sources \citep{https://doi.org/10.3847/1538-4357/aca524, https://doi.org/10.1093/mnras/stad203}, at least for the cooler sides of the companions away from the pulsars.

\section{Profile Evolution} \label{sec:Profile Evolution}

\begin{figure}[ht!]
\begin{center}
\includegraphics[width=0.5\textwidth,angle=0]{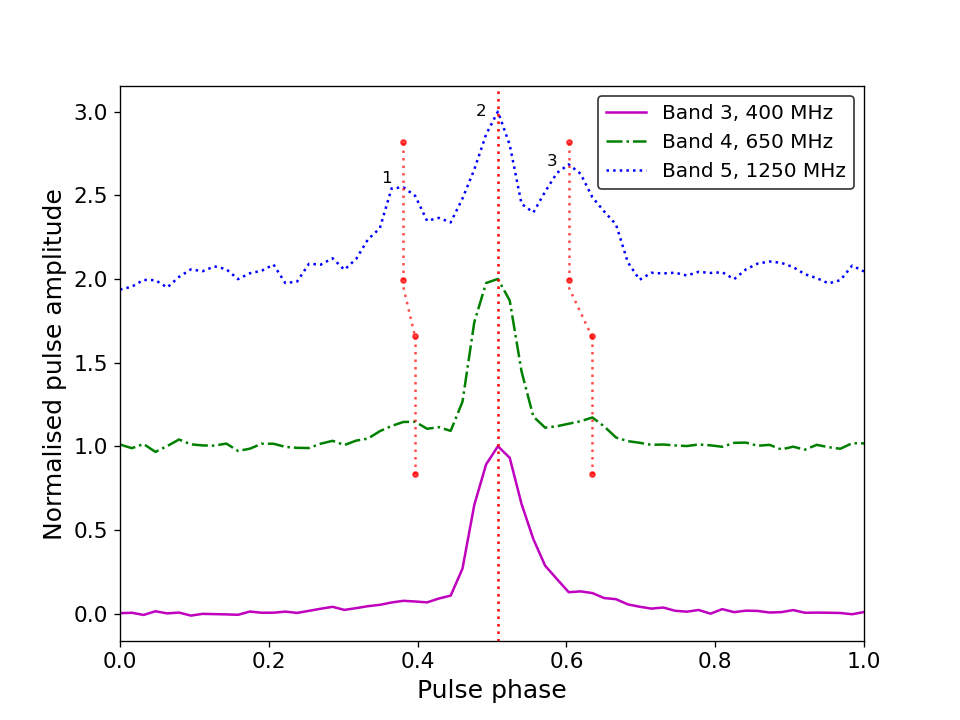}
\caption{Normalized averaged pulse profiles from coherently de-dispersed GMRT observation at 400 MHz (dotted blue line) and 650 MHz (dash-dotted green line), plus incoherently de-dispersed phased array observations at 1250 MHz (solid magenta line). The vertical dashed line is a visual guide to show how the main peaks at various frequencies are intrinsically aligned. The inflections in the red dotted line indicate shifts in the relative position of the components at higher frequencies.}
\label{fig:2}
\end{center}
\end{figure}

The frequency evolution characteristics of radio pulse profiles provide crucial insights into understanding the radiation mechanism of pulsars. Analyzing multiband radio pulse profiles can unveil various emission geometries of pulsars, including emission region heights, beam shapes, radius-to-frequency mapping (RFM), and more \citep{ https://doi.org/10.1093/mnras/234.3.477, https://dx.doi.org/10.1086/342136}. In our observations of MSP J1242$-$4712 in the different frequency bands of uGMRT, we noticed distinct changes in the pulse profile and its frequency dependence (Figure \ref{fig:2}). At lower frequencies, a relatively strong single component is evident, while higher frequencies show the presence of two weaker components along with the main component. Around 650 MHz, the pulsar exhibits a clear three-component pulse profile with the main component being the strongest. The main component consistently shows a measured $W_{50}$ of approximately 0.47 ms across all observed frequencies, whereas the $W_{50}$ of the other two components increases with frequency. Specifically, at 650 MHz and 1.2 GHz, the $W_{50}$ for peak 1 is 0.56 ms and 0.74 ms; for peak 3, it is 0.58 ms and 0.70 ms.

The interstellar medium (ISM) is a turbulent, ionized plasma that introduces time delays and phase variations in radio signals. This causes observable phenomena such as dispersion, scattering, angular broadening, and interstellar scintillation. Dispersion leads to a time delay between pulses at different frequencies in a broadband pulsar signal, while scattering results in temporal broadening, causing average profiles to appear asymmetrically broader at lower frequencies. To explain the pulse-broadening due to scattering, the thin screen model \citep{https://dx.doi.org/10.1086/338324, https://doi.org/10.1051/0004-6361:20035881} is assumed, and it can be described by an exponential function $\sim exp(-t/\tau_{sc})$, where $\tau_{sc}$ represents the scattering time scale \citep{https://doi.org/10.1038/218920a0}. The simple picture of multipath propagation reproduces the observed pulse scattering features and predicts a strong dependence on frequency and dispersion measure of the pulsar ($\tau_{sc}\propto \nu^{-4}DM^{2}$) under the assumption of Gaussian inhomogeneity in the electron density fluctuations in the ISM model \citep{1971ApJ...164..249L}.

From the profile evolution shown in Figure \ref{fig:2} for PSR J1242$-$4712, it is evident that significant scattering is present at lower frequencies. The scattering time scale is calculated to be $\tau$ = 0.45 ms at 320 MHz by fitting the pulse profile with a convolution of Gaussian and exponential functions assuming the thin screen model. We calculated this scattering time scale to be 0.39 ms using the DM model given by \cite{2004ApJ...605..759B}.

\section {Discussion}\label{sec:Discussion}

\subsection{PSR $J1242-4712$ as a ``spider MSP"}\label{sec:spider}

\begin{figure}[ht!]
\begin{center}
\includegraphics[width=0.5\textwidth,angle=0]{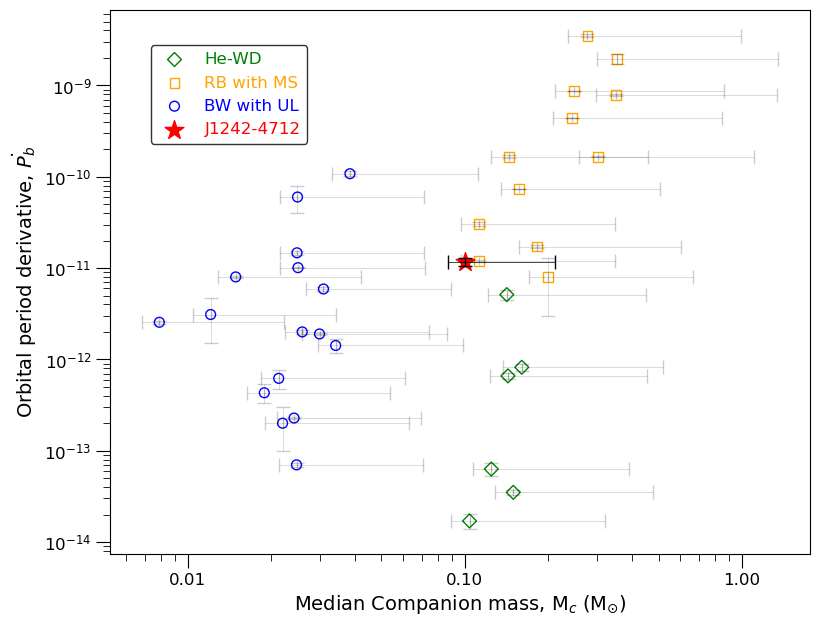}
\caption{Position of MSP J1242$-$4712 in $\dot{P_{b}}-M_{c}$ plot for  MSP binaries\textsuperscript{\ref{note2}} with UL, MS and He-WD companions. The companion masses are “median” masses that
assumed a 1.4 M$_{\odot}$ neutron star and an inclination of 60°; the error bars are estimated by varying inclination $i$ from 18° to 90°.}
\label{fig:pbdot-mc}
\end{center}
\end{figure}

Although there has been a significant increase in spider MSPs in recent years, binary evolution models \citep[e.g.,][]{2013ApJ...775...27C,2014ApJ...786L...7B} indicate that their fate differs from those of the He$-$WD (Helium-White Dwarf) MSP binaries that constitute the majority of binary MSP population.
Unlike the redbacks, systems featuring companions in the end stages of transitioning into He$-$WDs typically do not fill a substantial portion of their Roche lobes. This is because, during the MSP spin-up process, most of the companion's envelope has either been accreted or ejected \citep{2002ApJ...565.1107P}. Consequently, He$-$WD systems tend not to exhibit pronounced optical variability or radio eclipses. However, MSPs with non-degenerate stellar companions usually have been seen to exhibit radio eclipses and orbital period variations along with periodic optical variations due to tidal distortions of the secondary and irradiative heating of the tidally locked companion surface  \citep[e.g.,][]{https://doi.org/10.3847/1538-4357/abe1be, https://doi.org/10.3847/1538-4357/ac4ae4}.

Here we compare the behavior of the orbital-period derivative($\dot{P_{b}}$) in MSP binaries with $P_{b}<1$ day and $M_{c}<0.9$$M_{\odot}$ across various companion types in Figure \ref{fig:pbdot-mc}. We plotted the absolute values of orbital-period derivatives of such systems with UL(Ultra-light), MS(main sequence), and He-WD companions. We found nearly 20 such MSP binaries with He$-$WD companions in ATNF\textsuperscript{\ref{note2}}\citep{2005AJ....129.1993M}. However, very few such systems have $\dot{P_{b}}$ measurements. We found only six such systems, three in the Galactic field (J0751$+$1807, J1012$+$5307, and J1738$+$0333) and the rest in Globular clusters (J0024$-$7204U, J0024$-$7204Y, and J1701$-$3006B).
The MSP binaries with He$-$WD in Galactic field have $\dot{P_{b}}$$\sim$$10^{-14}$, which is two orders of magnitude lower than that of binaries in Globuler clusters, $\dot{P_{b}}$ $\sim$$10^{-12}$.
The $\dot{P_{b}}$ for Galactic field He$-$WDs is mostly contributed by the orbital decay from the emission of gravitational waves predicted by general relativity along with  Galactic acceleration and Shklovskii effect. For example $\dot{P_{b}}^{GR}\sim$ $-27.7^{+1.5}_{-1.9} \times 10^{-15}$ s/s for MSP J1738$+$0333, \citep{https://doi.org/10.48550/arXiv.1205.1450}.  Whereas the Globular cluster binaries will have a significant contribution to observed $\dot{P_{b}}$ from acceleration in the cluster \citep{1970SvA....13..562S,1991ApJ...366..501D}. Notably, a subtle correlation emerges between the observed $\dot{P_{b}}$ and companion's mass and type, that Galactic MSP binaries with a non-degenerate main sequence companion have a larger value of period derivative compared to He$-$WD companions for nearly the same companion mass because of the contribution of the tidal dissipation to the orbital decay $\dot{P_{b}}$ \citep{https://doi.org/10.1051/0004-6361/201937415}. 

The majority of redback systems show significant variations in their orbital period showing higher-order orbital period derivatives \citep{http://dx.doi.org/10.1088/0004-637X/807/1/18, http://dx.doi.org/10.3847/0004-637X/823/2/105}. As shown in Figure \ref{fig:pbdot-mc}, \psr\ exhibits substantial orbital period variability. Nevertheless, we could detect only up to the first order of orbital period derivative for this system for a timing baseline of 4.5 years (Figure \ref{fig:1}). The fairly large orbital period derivative can indicate that this may be a redback pulsar, and increasing the timing baseline by adding more sensitive observations may detect small variations of orbit over time and thereby higher-order orbital period derivatives for this system. While the detection of higher-order derivatives of the orbital period is common in the majority of redbacks, such derivatives have not yet been reported in certain redback systems, such as \psr\ (e.g. PSR J1023$+$0038 \citep{2016ApJ...830..122J, https://doi.org/10.3847/1538-4357/acee67}, PSR J2215$+$5135 \citep{2013ApJ...769..108B, https://doi.org/10.3847/1538-4357/acee67},  PSR J2129$-$0429 \citep{2018ApJ...861...89A,http://dx.doi.org/10.3847/0004-637X/823/2/105}, PSR J1908$+$2105 \citep{2021ApJ...909....6D}). However, PSRs J1023$+$0038 \citep{2016ApJ...830..122J},  J2215$+$5135, and J2129$-$0429 show significant residual orbital variations \citep{2024arXiv240209366B} which are too complex to be modeled. Therefore, the ephemeris provided by \citet{https://doi.org/10.3847/1538-4357/acee67} does not include higher-order orbital period derivatives. Several mechanisms such as changes in the companion star’s gravitational quadrupole moment \citep{1987ApJ...322L..99A, https://ui.adsabs.harvard.edu/link_gateway/1994ApJ...436..312A/doi:10.1086/174906} and asynchronous rotation of the companion star can give rise to tidal forces that transfer angular momentum from the star to the orbit  \citep{http://dx.doi.org/10.3847/2041-8213/833/1/L12}. These can elucidate the intricate orbital period variation over time in spider binaries as explained in \citet{https://doi.org/10.1093/mnras/staa3484}. For PSR J1242$-$4712, with long-term sensitive timing, it is plausible that we may eventually comprehend the orbital variations by modeling these effects. Moreover, we also observed eclipses in \psr\ (Section \ref{sec:Eclipse}). In light of this, it seems that \psr\ is a spider system with a semi- or non-degenerate companion. Yet, it is imperative to conduct follow-up optical observations to conclusively confirm this classification.

The timing solution also revealed that \psr\ is in a binary orbit with an orbital period of 7.7 hours and a minimum companion mass of 0.087 M$_{\odot}$. Figure \ref{fig:3} presents $P_{b}-M_{c}$ plot for all spider MSPs where estimates were available\textsuperscript{\ref{note2}}. PSR J1242$-$4712 occupies a relatively unexplored part in the $P_{b}-M_{c}$ plot, lying in an ambiguous region between typical black widows and redbacks. Thus, this MSP may aid in probing whether redbacks and black widows represent separate populations evolving independently due to bimodal evaporation efficiency \citep{2013ApJ...775...27C}, or if black widow systems represent a final stage in the evolution of certain redback systems \citep{2014ApJ...786L...7B}.\par

Following \citet{1999A&A...350..928T}, {\citet{https://doi.org/10.3847/1538-4357/aad5ec} computed correlation between $P_{b}$ and $M_{c}$ numerically for MSP binaries with WD companions and short orbital period ($P_{b}$ $\sim 1$ days) using the model,
\begin{equation}
 \frac{M_{c}}{M_{\odot}}=\left ( \frac{P_{b}}{b} \right )^{1/a} +c    \label{eq:1} 
\end{equation}
given by, \citet{1999A&A...350..928T}; where $P_{b}$ is in the unit of days. 
They reported fitted parameters $a$=$4.91 \pm 2.26$, $b$=$(4.18 \pm 13.1) \times 10^{5}$, $c$=$0.12 \pm 0.04$ for their sample MSP binaries which align with the findings of \citet{1999A&A...350..928T}. The dotted line in Figure \ref{fig:3} represents Equation \ref{eq:1} and indicates that the mass distribution in redback populations closely matches 
the model. Although \psr\ follows this model within the error bars it predicts a heavier companion mass indicating that it is more likely to be nearly face-on, which may also explain the observed short-duration eclipses for this system (Section \ref{sec:Eclipse}). However, if \psr\ is observed nearly face-on implying a more massive companion ($\sim$ 0.21 M$_{\odot}$), explaining optical non-detection (Section \ref{sec:X-ray and optical}) becomes more challenging. 

While many redback systems have bright optical counterparts (Figure \ref{fig:limits}), there are a number of redback systems, such as PSR J1908$+$2105 (M$_{c}$ $\sim$ 0.055 M$_{\odot}$), PSR J1957$+$2516 (M$_{c}$ $\ge$ 0.1 M$_{\odot}$), PSR J1302$-$3258 (M$_{c}$ $\ge$ 0.15 M$_{\odot}$ ) for which no optical counterparts are yet identified \citep{2019ApJ...872...42S}. If these systems are canonical redbacks, higher distance, and reddening could have led to optical non-detections, but the full range of effective temperatures for these systems has not been fully explored. The recent work by \citet{2024arXiv240209366B}, suggests that J1302$-$3258 shows occasional eclipsing at low frequencies for approximately 10\% of its orbit. Interestingly, despite this behavior, the companion mass aligns with that of a white dwarf, and it lacks other defining characteristics typically associated with redback systems. Also, \citet{2021ApJ...909....6D} characterized J1908$+$2105 as an intermediate case between black-widow and redback systems. This binary system exhibits an orbital period of 3.5 hours and a minimum companion mass of 0.055 solar masses, closely resembling black widows. However, it also displays extensive eclipses at lower frequencies, covering nearly 40\% of its orbit, very similar to the behavior observed in redback systems. Here to note, this system is also an example of a redback without higher-order orbital period derivatives.\par

The measured $\mathrm{\dot{E}}$ for \psr\ lies well within the values of $\dot{E}$ for MSPs ($\sim$ $10^{32}$ $-$ $10^{36}$ erg s$^{-1}$) listed in ATNF Pulsar catalog and is close to the values of $\dot{E}$ ($\sim$ $10^{34}$ erg s$^{-1}$) typically found for all Galactic redback MSPs \citep{https://dx.doi.org/10.1088/2041-8205/756/2/L25, https://ui.adsabs.harvard.edu/link_gateway/2013IAUS..291..127R/ADS_PDF, 2020ApJ...900..194K}.
Similarly, the pulsar wind flux at the companion ($\dot{E}/a^{2}$) is found to be $0.29 \times 10^{35}$ erg s$^{-1} R_{\odot}^{-2}$, which is comparable to what is typically found in spider MSPs \citep{2020ApJ...900..194K}. 

\begin{figure}[ht!]
\begin{center}
\includegraphics[width=0.5\textwidth,angle=0]{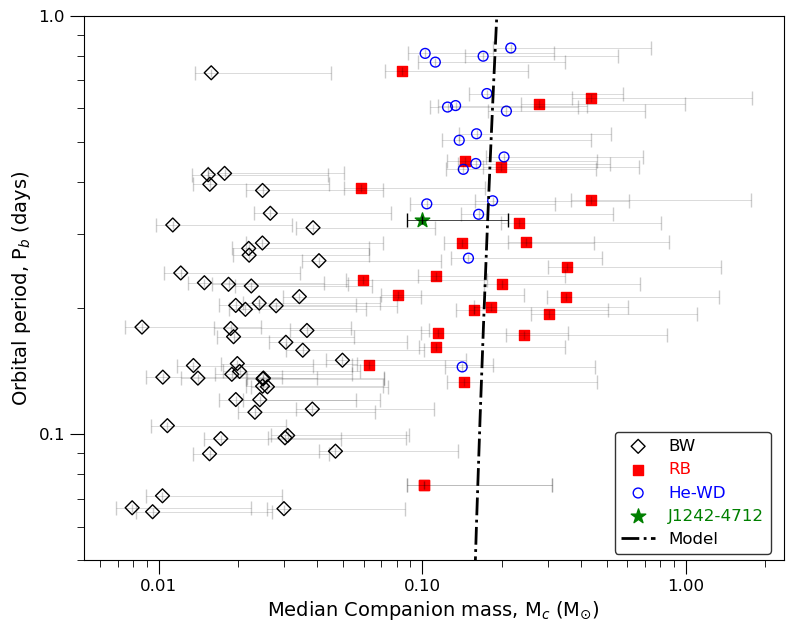}
\caption{Position of MSP J1242$-$4712 in $P_{b}-M_{c}$ plot for spider and He$-$WD MSP binaries\textsuperscript{\ref{note2}}. The black dotted line shows the best fit of the model (Equation \ref{eq:1}). The companion masses are ``median'' masses that assumed a 1.4 M$_{\odot}$ neutron star and an inclination of 60°; the error bars are estimated by varying inclination $i$ from 18° to 90°.}
\label{fig:3}
\end{center}
\end{figure}

\subsection{Possible eclipsing in the system}\label{sec:Eclipse}

\begin{figure}[ht!]
\begin{center}
\includegraphics[width=0.4\textwidth,height=0.40\textwidth,angle=0]{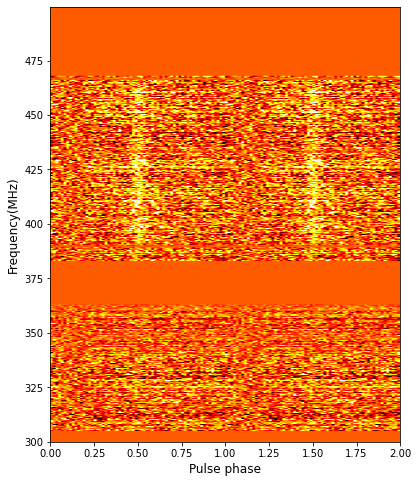}
\caption{uGMRT band-3 observations of J1242$-$4712; eclipse near superior conjunction of the pulsar (orbital phase $\sim$ 0.23-0.25) below 360 MHz.}
\label{fig:main_eclipse}
\end{center}
\end{figure}

Material blown from the companion may cause an eclipse of the pulsar's radio emission. The majority of spider MSP systems exhibit frequency-dependent eclipses, which appear more pronounced at lower frequencies than higher ones, often absent above $\sim$1.5 GHz  \citep{https://doi.org/10.1093/mnras/staa596, 2020ApJ...900..194K}. 
 For the majority of redbacks, eclipses near superior conjunction are observed for longer duration. At lower frequencies, eclipse fraction can even be $>$50\% of orbital phase   \citep{https://doi.org/10.3847/1538-4357/abbdfb}. With recent uGMRT observations of PSR J1242$-$4712 in phased array mode covering the eclipse phase, we have seen frequency dependent eclipse below 360 MHz, for a very short duration near superior conjunction of pulsar between orbital phase 0.23$-$0.25 (Figure \ref{fig:main_eclipse}). The other epochs covering the eclipse phase were observed using incoherent array mode, where large TOA uncertainties and a rather coarse sampling of the orbit could have masked the signature of eclipsing. The absence of long-duration low-frequency eclipses can constrain inclination angles and companion mass. In addition to the eclipse near the superior conjunction of the pulsar, we have also seen a mini eclipse near orbital phase 0.18 below 360 MHz.
Moreover, a clear signature of eclipsing is also observed for this MSP near the pulsar's inferior conjunction in band-4 (orbital phase $\sim$ 0.7, Figure \ref{fig:4}) in one of the epochs causing an excess DM delay and large error in pulses time of arrivals (TOAs) as also seen for redback MSP J1227$-$4853 \citep{2020ApJ...900..194K}. The observed mini eclipses indicate temporal clumping of the materials throughout the binary orbit which has also been observed in other such systems \citep{http://dx.doi.org/10.3847/0004-637X/823/2/105}.
The ongoing effort of simultaneous multifrequency observations in the phased array mode of the uGMRT will provide a better probe to possible eclipsing in this MSP. 

\begin{figure}[ht!]
\begin{center}
\includegraphics[width=0.4\textwidth,height=0.45\textwidth,angle=0]{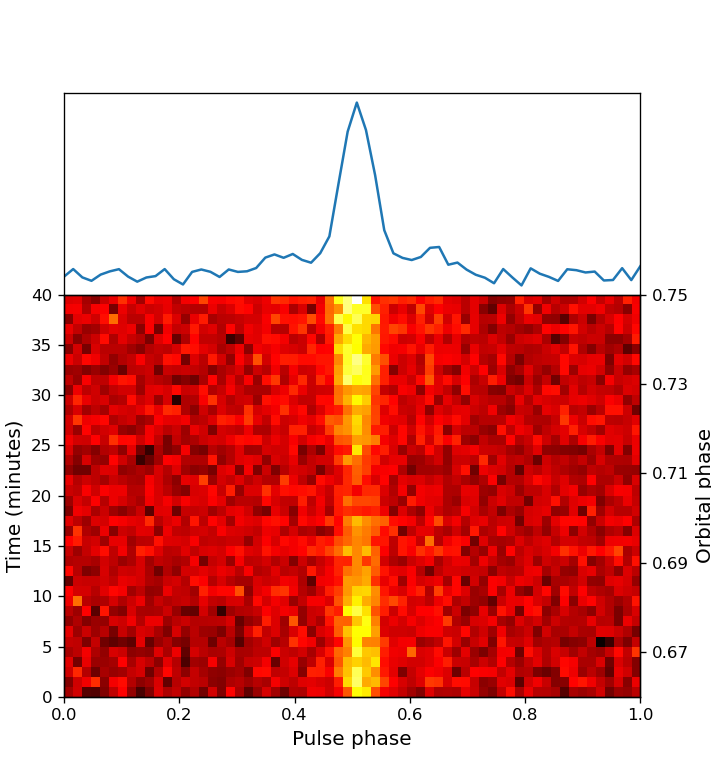}
\caption{uGMRT band-4 observations of J1242$-$4712; indication of partial eclipsing around orbital phase 0.68$-$0.73 (MJD 59835).}
\label{fig:4}
\end{center}
\end{figure}

\subsection{Possible inclusion of PSR $J1242-4712$ in PTAs}\label{sec:PTA}

The Pulsar Timing Array (PTA, e.g.\citet{https://ui.adsabs.harvard.edu/link_gateway/1979ApJ...234.1100D/doi:10.1086/157593}) employs a collection of MSPs with precise timing parameters to detect Gravitational wave (GW) signals through the angular correlation between the residuals of the arrival timings of pairs of MSPs. Adding more MSPs to the PTAs aids in the faster detection of GW signals with increased sensitivity and resolution.
The recent timing of PSR J1242$-$4712 shows a timing residual of 2.3 $\mu$s, achieved while using the sensitive coherently dedispersed phased array data from uGMRT observations having 200 MHz bandwidth from 2022 to 2023 (Figure \ref{fig:1}). This timing precision is similar to the timing precision of spider MSPs, PSR J1802$−$2124 \citep[$\sim$ 2.3 $\mu$s,][]{http://dx.doi.org/10.1088/0004-637X/711/2/764}, PSR J1012$+$5307 \citep[$\sim$ 3.1 $\mu$s,][]{https://ui.adsabs.harvard.edu/link_gateway/2001MNRAS.326..274L/ADS_PDF},  and lower than the timing precision of PSR J2214+3000 \citep[$\sim$ 4.3 $\mu$s,][]{2020MNRAS.494.2591B}, which have already been added to the European pulsar timing array \citep{https://doi.org/10.1093/mnras/stw347}. The present precision timing using only PA and CD observations and allowing TOAs with errors less than 10 $\mu$s produced a weighted timing residual of 2.4 $\mu$s (Figure \ref{fig:1}), but more sensitive observations that allow TOAs to have even smaller errors may yield a more precise timing model with a smaller rms. Also, the ongoing observations with the uGMRT will allow us to add additional sensitive simultaneous dual frequency TOAs, which may enable better modeling of the PSR J1242$-$4712 with an extended timing baseline, increasing its possibility of inclusion in PTAs. 

\section{Conclusion} \label{sec:Conclusion}
In this paper we present a study of \psr\, having a periodicity of 5.31 ms and a dispersion measure (DM) of 78.6 pc cm$^{-3}$, discovered by the uGMRT. Observations using uGMRT at band-3 and band-4 revealed that \psr\ shows significant changes in the pulse profile with frequency, showcasing scattering effects. At lower frequencies, a relatively strong single component is evident, while at 650 MHz, J1242$-$4712 displays a three-component profile. Profiling the scattering time scale at 320 MHz yields $\tau$ = 0.45 ms. 

The timing solution for J1242$-$4712, achieved using the uGMRT band-3 and band-4 observations spanning the 4.5 years from 2019 to 2023, indicates that it is in a 7.7-hour binary orbit with a companion having a median mass of 0.1 M$_{\odot}$. This MSP is located in a relatively unexplored region in the orbital period versus companion mass space between black widows and redbacks for such compact orbit spider MSP systems.
We expect to observe complete radio eclipses around the pulsar's superior conjunction in redback systems, where the radio flux disappears entirely. This is because the companion stars have stronger winds that result in an intra-binary shock that wraps around the pulsar, leading to radio eclipses over a large fraction of the orbit \citep{https://doi.org/10.3847/1538-4357/abbdfb}. However, PSR J1242$-$4712 exhibits a very short duration of eclipse near superior conjunction of the pulsar (orbital phase $\sim$ 0.23$-$0.25) along with a couple of mini eclipses in other orbital phases.
\psr\ also exhibits substantial orbital period variability, even though we could detect only up to the first order of orbital period derivative. The observed orbital period variability and eclipses suggest that this is a ``spider" MSP binary containing a semi- or non-degenerate companion, not a He$-$WD system.  We could not yet identify an optical counterpart for this system (Section \ref{sec:X-ray and optical}), which is observed for the majority of other redback systems. This optical non-detection could be attributed to reddening due to distance.

Positioned within an ambiguous region between the conventional black widow and redback characteristics, this system emerges as a noteworthy and unusual redback variant sharing properties of both black widows and redbacks. These findings suggest a category of objects that share properties bridging two subclasses of spider binary systems.

\par

We acknowledge the support of the Department of Atomic Energy, Government of India, under project No.12-R\&D-TFR5.02-0700. The GMRT is run by the National Centre for Radio Astrophysics of the Tata Institute of Fundamental Research, India. We acknowledge the support of GMRT telescope operators for observations. D.L.K. was supported by the NANOGrav NSF Physics Frontiers Center award numbers 1430284 and 2020265. 
Portions of this work performed at NRL were supported by NASA. We thank the anonymous referee for insightful suggestions that contributed to the improvements of the paper.

\bibliography{J1242-4712}{}
\bibliographystyle{aasjournal}

\end{document}